\documentclass[a4paper,
               biblatex,     % biblatex is used
               ]{jacow}
\usepackage{subfig}

\makeatletter%
	\ifboolexpr{bool{xetex}}
	 {\renewcommand{\Gin@extensions}{.pdf,%
	                    .png,.jpg,.bmp,.pict,.tif,.psd,.mac,.sga,.tga,.gif,%
	                    .eps,.ps,%
	                    }}{}
\makeatother

\ifboolexpr{bool{xetex} or bool{luatex}} % test for XeTeX/LuaTeX
 {}                                      % input encoding is utf8 by default
 {\usepackage[utf8]{inputenc}}           % switch to utf8

\usepackage[USenglish]{babel}

\ifboolexpr{bool{jacowbiblatex}}%
 {%
  \addbibresource{refs.bib}
 }{}
\begin{document}

\title{Recent advances in \\metallographic polishing for SRF application}

\author{O. Hryhorenko\thanks{hryhoren@jlab.org}, Thomas Jefferson National Accelerator Facility (JLAB), Newport News, VA, USA\\
        C.~Z.~Antoine, T.~Proslier, F. ~Eozenou, Université Paris-Saclay, CEA Département des Accélérateurs\\ de la Cryogénie et du Magnétisme (CEA-IRFU), Gif-sur-Yvette, France \\
        T. Dohmae, High Energy Accelerator Research Organization (KEK), Tsukuba, Japan\\
        S.~Keckert, O.~Kugeler, J.~Knobloch, Helmholtz Zentrum Berlin (HZB), Berlin, Germany\\
        D.~Longuevergne, Université Paris-Saclay, CNRS/IN2P3, IJCLab, Orsay, France}

\maketitle

\begin{abstract}
This paper is an overview of the metallographic polishing R\&D program covering Niobium and Copper substrates treatment for thin film coating as an alternative fabrication pathway for 1.3 GHz elliptical cavities. The presented research is the result of a collaborative effort between IJCLab, CEA/Irfu, HZB, and KEK in order to develop innovative surface processing and cavity fabrication protocols capable of meeting stringent requirements for SRF surfaces, including the reduction of safety risks and ecological footprint, enhancing reliability, improving the surface roughness, and potentially allowing cost reduction. The research findings will be disclosed.
\end{abstract}

\section{INTRODUCTION}

The SRF cavities made of bulk Niobium are reaching their theoretical performance limits via different surface processing techniques (surface roughness decrease, interstitials engineering,...) \cite{doi:10.1080/21663831.2022.2126737, 10.1093/ptep/ptab056, Bafia:2019fdi}. The global strategy for future decades in the field of superconducting radiofrequency (SRF) technology is focused on the fabrication of novel superconducting cavities made of alternative superconductors and multilayer structures \cite{valentefeliciano2022nextgeneration}. This approach aims to enhance performance beyond the theoretical limits of bulk Nb and improve the accelerator cost-efficiency using thin films with the ultimate goal of operation at 4.2K. A key aspect of this strategy involves optimizing and industrializing the substrate preparation (cavity), specifically focusing on materials such as Nb and Cu. This paper examines the protocols involved in the pursuit of substrate preparation via metallographic polishing (MP) technology for niobium and copper, highlighting the potential benefits of MP in the preparation of flat samples. In the framework of the IFAST project, RF disks and Quadrupole (QPR) resonator samples are used for superconducting RF evaluation of deposited thin films \cite{antoine:hal-04021253}. 

Samples were provided by STFC (Daresbury, UK) and HZB (Berlin, Germany) labs, were polished at IJCLab, and their surface quality was evaluated. The disclosed findings presented here contribute to the ongoing efforts in investigating an alternative cavity fabrication pathway, see related conference paper\cite{hryhorenko:srf23}, in the framework of the FJPPL program. Moreover, in this work we show the first successful results of the RF test achieved on the QPR sample after metallographic polishing. 

\section{Experimental Set-up and \\Quality Control}

RF disks and QPR samples were polished with a commercial metallographic polishing (MP) machine MASTERLAM 1.0 (LAM PLAN production) at IJCLab. The surfaces were studied and the achieved quality was controlled after each manipulation via visual inspection and roughness measurements using a laser confocal microscope Keyence VK-X 200. The roughness measurements were performed on 9 different spots with a scan area 290 µm X 215 µm. Two parameters as average roughness (Sa), and maximal height deviation (Sz) were monitored and recorded. To complete characterization of the MP influence on the SRF surface at a high magnetic field, a QPR sample was evaluated under RF at HZB.

\subsection{RF Disks: Copper and Niobium Processing}

\begin{figure}
   \centering
   \includegraphics*[width=.7\columnwidth]{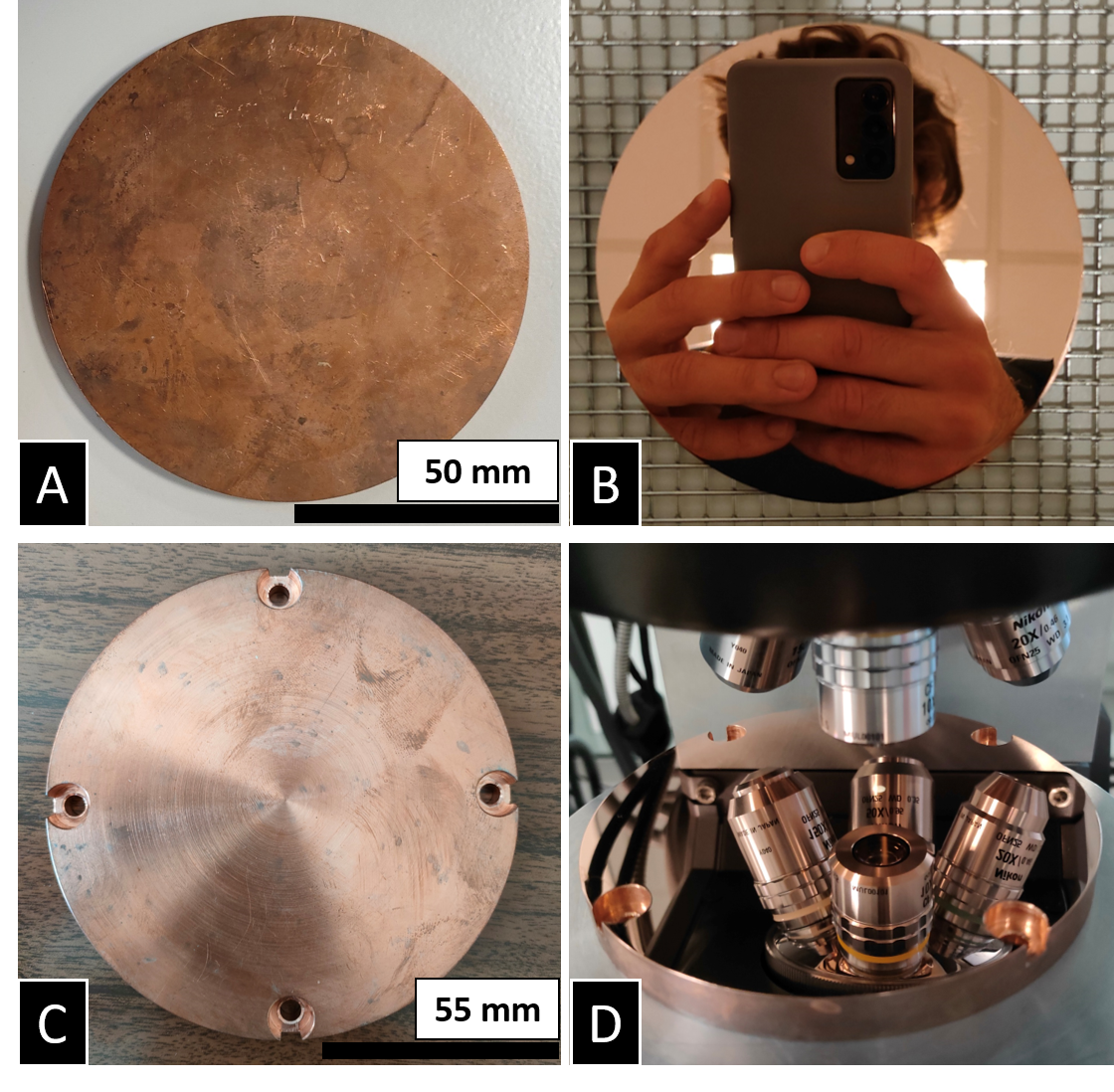}
   \caption{Photographs of the Copper RF disks used at STFC for RF characterization at 7.8 GHz: a) design 1 before MP polishing, b) design 1 after MP polishing, c) design 2 before MP, d) design 2 after MP. Note: samples are used in two different designs. Design 1 (diameter of 100 mm) requires the indium brazing between the sample and the support holder. Design 2 uses bolts (diameter of 110 mm). }
   \label{fig:Cu-disk}
\end{figure}

Disks of the two types, see Fig.~\ref{fig:Cu-disk}, are utilized at STFC for RF measurements with a choke cavity operated at 7.8 GHz, at a temperature of 4\,K, and with surface magnetic fields up to 1 mT \cite{seal2022rf}. Type 1 (diameter of 100 mm), which can be made of copper (Cu) or Niobium (Nb), requires indium brazing between the sample and support holder, so indium removal after characterization is required. Type 2 (diameter of 110 mm), which is exclusively applied to copper, is based on a slightly complicated design as bolts are used to connect the sample with a choke cavity, resulting in the surface being drilled, but the step with indium removal is omitted. 

The 5-step procedure described in Table \ref{tab:Cu_recipe} was used to MP polish  RF disks made of copper. The recipe was applied on 3 disks of Type 1 and on one disk of Type 2. The first three steps are focused to planarize the surface. The last 2 steps are focused on removing the damage from the previous steps and creating a mirror-smooth and scratch-free surface. Considering prior experience with Nb (niobium), it was determined that the 5-step procedure could be replaced with a 2-step recipe \cite{hryhorenko:tel-02455975}; however, this modification would significantly prolong the preparation time.
\begin{table*}[!tbh]

	\setlength\tabcolsep{3.5pt}
	\centering
	\caption{Recipe for Cu Planar Samples}
	\label{tab:styles}
		\begin{small}
	\begin{tabular}{cccccc}
		\toprule
		\textbf{Step} & \textbf{1}               & \textbf{2}  & \textbf{3}   & \textbf{4} & \textbf{5} \\
		\midrule
  \textbf{Disk/Commercial name} & Resin/Platinum 1 & Resin/Platinum 4 & Resin+Cu/Cameo Gold & Woven fibre/3TL1 & Viscous fibre/4FV3 \\ 
  \midrule
		\textbf{Abrasive/Size, um} & diamond/125 & diamond/15 & diamond/3   & diamond/3 & diamond/1\\ 
  \midrule
  \textbf{Applied pressure, kPa } & 20 & 25  & 20-25   & 30 & 30 \\ 
        \midrule
        \textbf{Rotation speed, RPM} & 150/150 & 150/150  & 150/150   & 60/150 & 60/150 \\ 
        \midrule
        \textbf{Time, min} & Until plane & 15  & 15   & 40 & 10\\ 
		\bottomrule   %\SI{0.25}{in}
	\end{tabular}
	 	\end{small}
 \label{tab:Cu_recipe}
\end{table*}

 In addition to procedure steps, proper cleaning is required between steps. At the end of each step, the surface was rinsed with ultra-pure water, cleaned in the ultrasound bath (only de-ionized water), and dried with a Nitrogen gun. In order to improve the surface state, at the last polishing step, abrasives supplies are replaced by deionized water on the polishing pad, so superior cleaning is achieved. The time of cleaning varies depending on the sample size.

\begin{figure}
   \centering
   \includegraphics*[width=\columnwidth]{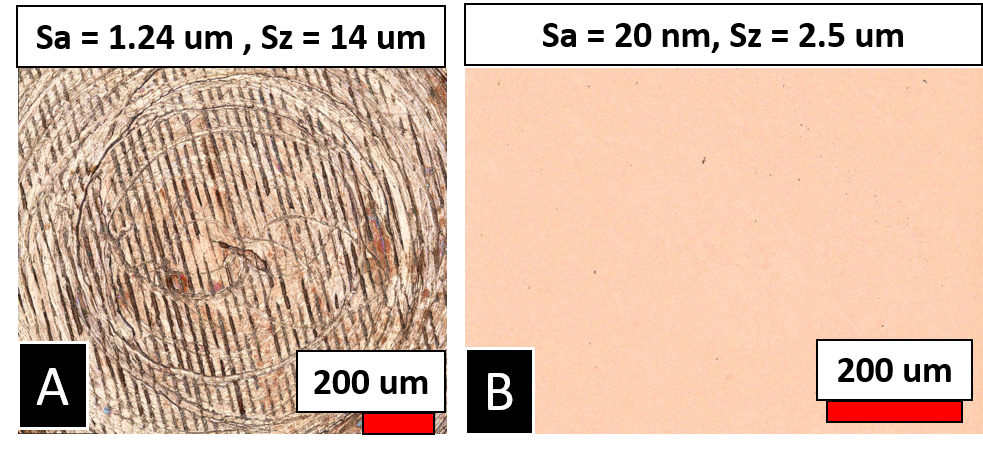}
   \caption{Visual inspection of Cu disks with a laser confocal microscope: a) after machining, b) after MP polishing.}
   \label{fig:cu-disks-before-after-polishing}
\end{figure}

As can be seen, in Fig.~\ref{fig:cu-disks-before-after-polishing}, the initial patterns after machining were removed by the MP technology in less than 90~minutes, and roughness parameters decreased from Sa=1.24\,µm, Sz=14\,µm down to Sa=0.02\,µm, Sz= 2.5\,µm.

\begin{figure}
   \centering
   \includegraphics*[width=.9\columnwidth]{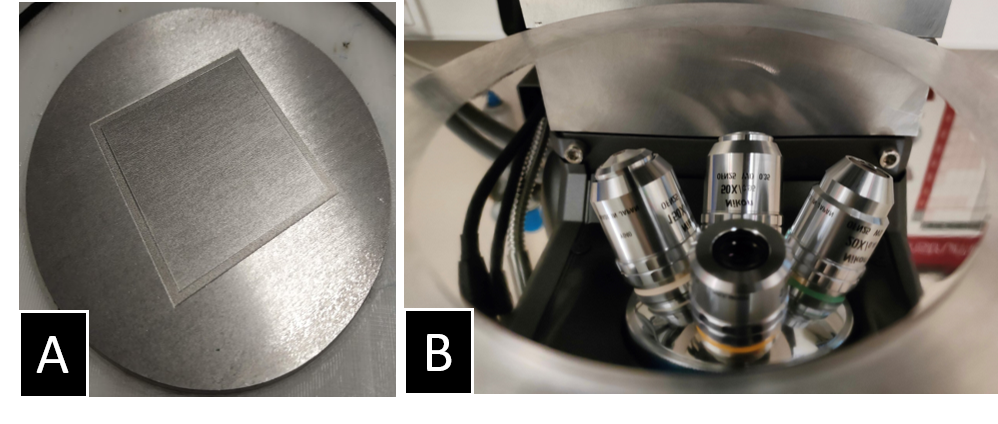}
   \caption{Photographs of the Nb RF disks used at STFC for RF characterization at 7.8 GHz: a) "as received", b) after MP.}
   \label{fig:Nb-disk}
\end{figure}

The niobium RF disk with a RRR of 400 underwent an MP procedure, see Fig.~\ref{fig:Nb-disk}. The received disk was found to have the following surface roughness parameters: \mbox{Sa = 2.2 ± 0.3 µm}, Sz = 21 ± 4 µm. High Sz indicates non-optimal flatness, so the sample undergoes pretreatment through a grinding process, a 200 µm layer removal. Subsequently, a standard 2-step procedure \cite{hryhorenko:srf19-thp002} was applied.

\begin{figure}[!tbh]
   \centering
   \includegraphics*[width=\columnwidth]{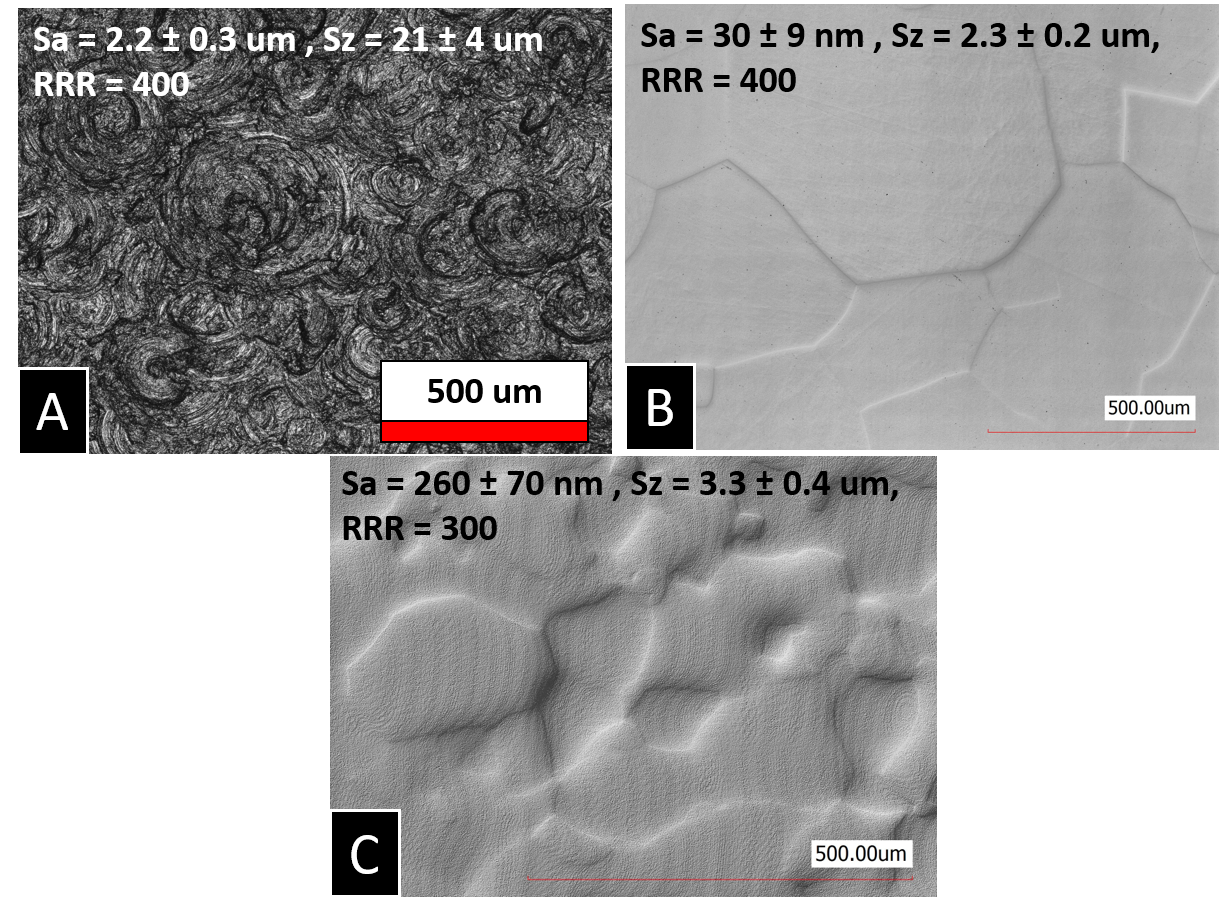}
   \caption{Visual inspection of Nb disks with a laser confocal microscope: a) "as received", b) after MP polishing (RRR=400), and c) after MP polishing (RRR=300).}
   \label{fig:nb-disks-before-after-polishing}
\end{figure}

The recipe is presented in Table ~\ref{tab:Nb_recipe}.  The obtained roughness are Sa = 30 ± 9 nm, Sz = 2.3 ± 0.2 µm. In cavity fabrication, it is typically common to use a material grade RRR=300. However, in the latter case, roughness is higher (typically Sa = 260 nm after 2 hours cycle) after an identical 2 steps polishing procedure, see the comparison in Fig.~\ref{fig:nb-disks-before-after-polishing}. 

\begin{table}[!tbh]
	\setlength\tabcolsep{3.5pt}
	\centering
	\caption{Recipe Developed at IJCLab/CEA-IRFU for Nb Planar Sample \cite{jmmp7020062}.}
	\label{tab:styles}
	\begin{small}
	\begin{tabular}{p{0.25\linewidth}  p{0.32\linewidth}  p{0.35\linewidth}}
		\toprule
		\textbf{Step} & \textbf{1}               & \textbf{2}  \\
		\midrule
  \textbf{Disk}  & Resin+Cu/Cameo Gold & Micro-porous polyurethane/4MP2 \\ 
  \midrule
		\textbf{Abrasive/Size} & diamond/3 µm   & SiO\textsubscript{2}/50 nm + H\textsubscript{2}O\textsubscript{2} + NH\textsubscript{4}OH diluted in water (20\%) \\ 
   \midrule
   \textbf{Applied pressure, kPa} & 10-15   & 15-20 \\ 
   \midrule
   \textbf{Rotation speed, RPM} & 150/150   & 150/150 \\ 
   \midrule
   \textbf{Time, min} & To remove damaged layer   & 480 \\ 
		\bottomrule   %\SI{0.25}{in}
	\end{tabular}
	\end{small}
 \label{tab:Nb_recipe}
\end{table}

\begin{figure}[!tbh]
   \centering
   \includegraphics*[width=\columnwidth]{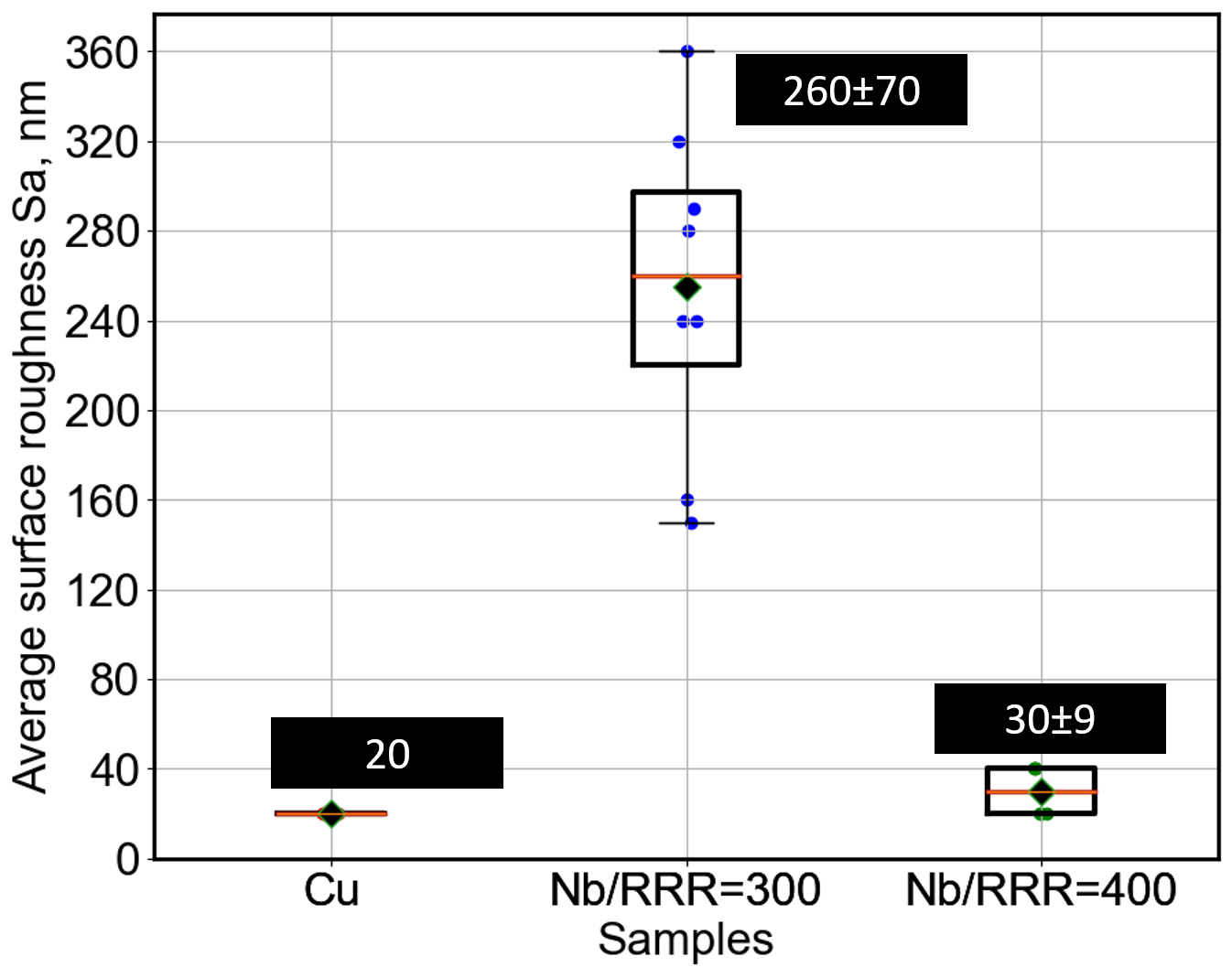}
   \caption{Average surface roughness Sa versus MP polished sample.}
   \label{fig:roughness_samples}
\end{figure}

As can be seen in Fig.~\ref{fig:roughness_samples}, the average roughness is compared for different material polished. Based on the visual inspection the revealed grains for RRR=400 grade are significantly bigger (>500 µm) compared to RRR=300 \mbox{(100-200\,µm).} 

Polished RF disks in this work will be used for thin-film deposition activities in the framework of the IFAST project at STFC. A Nb RF disk before and after MP polishing was evaluated using a choke cavity, and results were already published, see paper D. Seal \textit{et al.}~\cite{seal2022rf}. However in order to verify to assess the surface properties are close to the defined specifications (lower frequency, higher external magnetic field), an advanced characterization with a QPR system is required.\\

\subsection{QPR Niobium Processing}
\begin{figure}[!tbh]
   \centering
   \includegraphics*[width=\columnwidth]{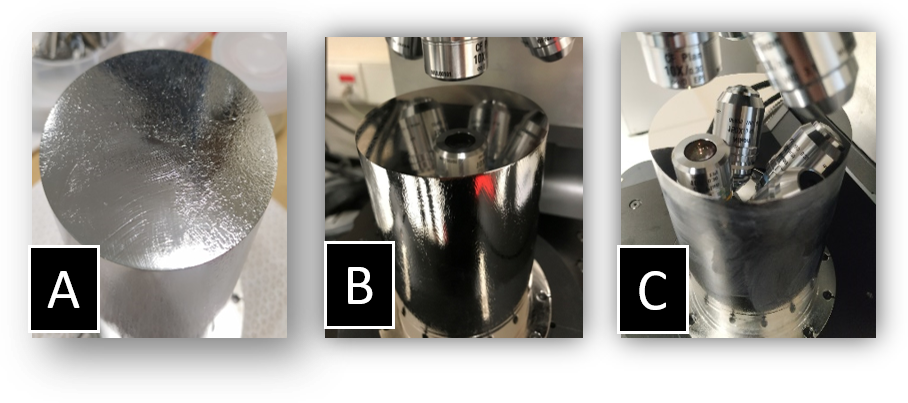}
   \caption{Visual inspection of the QPR sample: a) after BCP, b) after EP, c) after MP.}
   \label{fig:photo_inspection_qpr}
\end{figure}

In order to optimize the superconducting properties of the material, the QPR samples undergo surface treatment compatible with a cavity preparation. In addition, complete removal of In from the sample flange. Figure~\ref{fig:photo_inspection_qpr} shows a typical surface after BCP, EP and MP.

\begin{figure}[!tbh]
   \centering
   \includegraphics*[width=.7\columnwidth]{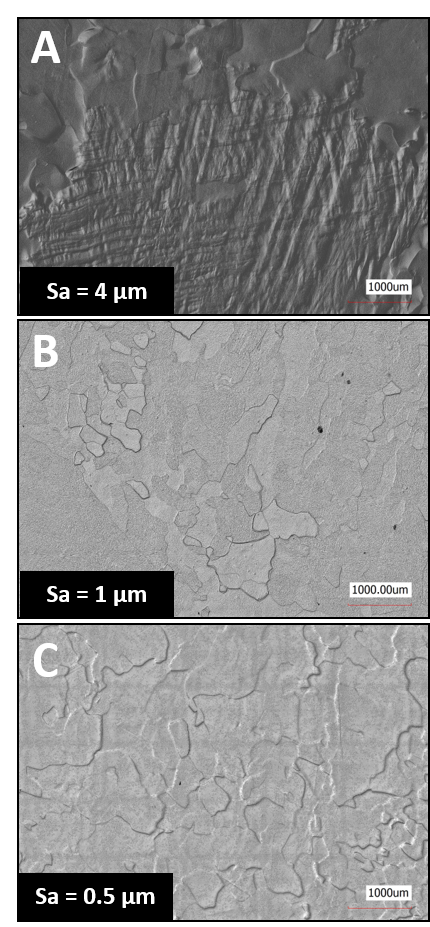}
   \caption{Images of the QPR sample taken with a laser confocal microscope: a) after BCP, b) after EP, c) after MP.}
   \label{fig:confocal_inspection_qpr}
\end{figure}

 The roughness was evaluated in accordance with to type of performed polishing, see Fig.~\ref{fig:confocal_inspection_qpr}. The MP-treated surface shows a superior surface (Sa=0.5 µm) compared to the conventional treatment (EP=1 µm). However, to validate this technology for substrate preparation, a RF test is required.

RF characterization was performed with a Quadrupole resonator (QPR) at HZB lab \cite{Keckert:2156188}. This technique gives a unique possibility to test flat samples by a calometrique method at three operation frequencies (415 MHz, 847 MHz and 1289 MHz) and at a high RF field \cite{10.1063/5.0046971}. The design of the system is based on the QPR samples with a diameter of 75 mm, and an Indium brazing is required as an intermediate step to mount the sample to the test bench.

The baseline measurement (QPR run \#32) history:
\begin{itemize}
    \item 100 µm removal by EP.
    \item  Annealing 900\,°C for 3\,hours.
    \item Flash 20\,µm removal by EP.
\end{itemize}

The MP measurement (QPR run \#38) history:
\begin{itemize}
    \item Step 1: 20\,µm removal by MP (abrasion step).
    \item Step 2: 5\,µm removal by MP (polishing step).
    \item  Annealing 600\,°C for 10\,hours.
\end{itemize}

\begin{figure}
   \centering
   \includegraphics*[width=\columnwidth]{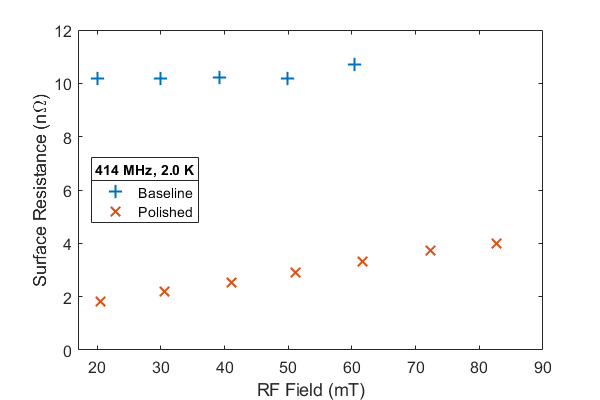}
   \caption{Surface resistance measurements towards RF field. Note: baseline is electropolished QPR, polished is MP processed.}
   \label{fig:Q1_RvsB_2K}
\end{figure}

Figure~\ref{fig:Q1_RvsB_2K} depicts measured data of surface resistance as a function of RF fields for samples after baseline and after MP polishing processing. The residual surface resistance was extracted from the measured surface resistance and resulted in residual resistance lower than 1 nOhm in the case of the MP polished surface, and lower than 10 nOhm in the case of EP treated surface. Improved surface preparation pushed the RF fields above 80 mT, as a result,  less incident power is dissipated.

\section{CONCLUSION}

We have shown that the preparation of the flat samples with the MP technology gives superior smoothness and flatness in comparison with the conventional treatment such as BCP or EP.  The baseline MP recipe can be used as a robust method for the production of the substrates with the following deposition of thin films in the framework of the IFAST project. Additional studies are still required to characterize the cavity performance with this technology, as the flat surface will be replaced with curvature. The following studies will be covered in accordance with the FJPPL program. 

\section{ACKNOWLEDGEMENTS}

The authors would like to mention that the MP polishing and visual inspection were done at Plateforme Vide \& Surface (IJCLab). Part of this work was supported by the European Nuclear Science and Application Research-2 (ENSAR-2) under grant agreement N° 654002. This project has received funding from the European Union’s Horizon 2020 Research and Innovation programme (IFAST) under Grant Agreement No 101004730 to perform an RF test. \mbox{O.~Hryhorenko} is currently supported by the U.S. Department of Energy, Office of Science, Office of Nuclear Physics under contract DE-AC05-06OR23177. 

\printbibliography

% for use as JACoW template the inclusion of the ANNEX parts have been commented out
% to generate the complete documentation please remove the "%" of the next two commands
% 
%%%\newpage

%%%\include{annexes-A4}

\end{document}